\def \SAIT #1 #2 {{\em Mem.\ Soc.\ Astron.\ It.\/} {\bf #1}, #2}
\def \MESS #1 #2 {{\em The Messenger\/} {\bf #1}, #2}
\def \ASTRNACH #1 #2 {{\em Astron. Nach.\/} {\bf #1}, #2}
\def \AAP #1 #2 {{\em Astron. Astrophys.\/} {\bf #1}, #2}
\def \AAL #1 #2 {{\em Astron. Astrophys. Lett.\/} {\bf #1}, L#2}
\def \AAR #1 #2 {{\em Astron. Astrophys. Rev.\/} {\bf #1}, #2}
\def \AAS #1 #2 {{\em Astron. Astrophys. Suppl. Ser.\/} {\bf #1}, #2}
\def \AJ #1 #2 {{\em Astron. J.\/} {\bf #1}, #2}
\def \ANNREV #1 #2 {{\em Ann. Rev. Astron. Astrophys.\/} {\bf #1}, #2}
\def \APJ #1 #2 {{\em Astrophys. J.\/} {\bf #1}, #2}
\def \APJL #1 #2 {{\em Astrophys. J. Lett.\/} {\bf #1}, L#2}
\def \APJS #1 #2 {{\em Astrophys. J. Suppl.\/} {\bf #1}, #2}
\def \APSS #1 #2 {{\em Astrophys. Space Sci.\/} {\bf #1}, #2}
\def \ASR #1 #2 {{\em Adv. Space Res.\/} {\bf #1}, #2}
\def \BAIC #1 #2 {{\em Bull. Astron. Inst. Czechosl.\/} {\bf #1}, #2}
\def \JSQRT #1 #2 {{\em J. Quant. Spectrosc. Radiat. Transfer\/} {\bf #1}, #2}
\def \MN #1 #2 {{\em Mon. Not. R. Astr. Soc.\/} {\bf #1}, #2}
\def \MEM #1 #2 {{\em Mem. R. Astr. Soc.\/} {\bf #1}, #2}
\def \PLR #1 #2 {{\em Phys. Lett. Rev.\/} {\bf #1}, #2}
\def \PASJ #1 #2 {{\em Publ. Astron. Soc. Japan\/} {\bf #1}, #2}
\def \PASP #1 #2 {{\em Publ. Astr. Soc. Pacific\/} {\bf #1}, #2}
\def \NAT #1 #2 {{\em Nature\/} {\bf #1}, #2}

\documentstyle[twoside]{memsait}
\input epsf.sty
\begin{opening}
\title{ASCA AND ROSAT X-RAY SPECTRA OF IC 5063 AND\\ NGC 3998}
\author{C. VIGNALI$^1$, A. COMASTRI$^2$, M. CAPPI$^3$, G.G.C. PALUMBO$^{1,3}$}
\institute{$^1$Dipartimento di Astronomia, 
Universit\`a di Bologna, via Zamboni 33, I-40126 Bologna,
Italy\\
$^2$Osservatorio Astronomico, via Zamboni 33, I-40126 Bologna, Italy\\
$^3$ITeSRE/CNR, via Gobetti 101, I-40129 Bologna, Italy}
\date{} 
\end{opening}

\begin{document}

\oddpagefooter{}{}{} 
\evenpagefooter{}{}{} 
\ 
\bigskip

\begin{abstract}
The results of the 0.1-10 keV spectral analysis of IC 5063 (Narrow 
Line Radio Galaxy) and NGC 3998 (LINER) are presented. In both cases 
there is evidence of a FeK$\alpha$ emission line, but its width is 
poorly constrained. 
The highly absorbed spectrum of IC 5063 is well 
described by a double power law model. The X-ray luminosity of a few $\times$ 
10$^{43}$ ergs s$^{-1}$ reveals a Sey 1 nucleus seen nearly edge-on, 
as confirmed by optical and IR studies. 
NGC 3998 is well fitted by a power law 
model ($\Gamma$$\sim$1.9) without reflection and weak absorption. 
The weak X-ray emission detected by ASCA and ROSAT suggests 
a low-luminosity AGN or an active nucleus in a low-accretion period.
\end{abstract}
%

\section{IC 5063}
IC 5063 is classified as a Narrow Line Radio Galaxy (z=0.011) 
(Caldweel \& Phillips 1981). 
IC 5063 was observed with the gas imaging spectrometer (GIS) and solid 
state spectrometer (SIS) on the ASCA satellite (Tanaka et al. 1994) 
in April 1994 and with the ROSAT PSPC (Pfeffermann et al. 1987) from 
November 1991 to April 1992. 
After applaying standard criteria to the data, about 19 Ks of useful 
data for ROSAT and $\sim$ 21(24) Ks for each SIS(GIS) detector were collected.\\
The combined analysis of ASCA and ROSAT data reveals a soft X-ray excess, a hard continuum 
and a reflection component. 
A double power law model plus a reflection component (fixed to 
cover 2$\pi$ steradiants at the source) provides a good fit to 
the ROSAT and ASCA data, but the spectral slope of the soft power law 
($\Gamma_{\rm S}$=2.16$^{+0.29}_{-0.31}$, errors at 90\%, 
$\Delta$$\chi^{2}$=2.71), steeper than the hard 
($\Gamma_{\rm H}$=1.70$^{+0.18}_{-0.19}$), indicates 
a more complex soft X-ray emission. 
A $L_{\rm soft}$/$L_{\rm hard}$ $\sim$1\% ratio is obtained by a
partial covering model. 
A bremsstrahlung model for the soft emission is equally good (see table 1). 
The broad H$\alpha$ line in polarized flux 
(FWZI$\sim$6000-9000 Km s$^{-1}$, Axon et al. 1994) and the anisotropy 
of the radiation field (``X'' geometry), with an opening angle of 
$\sim$50$^{\circ}$ (Colina et al. 1991), suggest that the observed soft 
X-ray component may be emission scattered along the line of sight from ionized 
gas above an absorbing torus. 
The column density value, $N_{\rm H}$=2.15$\pm{0.18}$ 
$\times$ 10$^{23}$ cm$^{-2}$, and the dust lane across the galaxy, likely 
responsible for the IR emission (Axon et al. 1982), agree with a nearly edge-on 
configuration. The 2-10 keV luminosity L$\sim$2$\times$10$^{43}$ ergs s$^{-1}$ 
indicates a Sey1-like active nucleus for IC 5063. 
Strong evidence for the FeK$\alpha$ 
line is given only by the SIS1 detector. The line parameters are given in 
table 2.\\

\centerline{\bf Tab. 1 - Spectral parameters}

\begin{table}[h]
\hspace{0.80cm} 
\begin{tabular}{|l|c|c|c|c|c|}
\hline
\hline
source&model&$\Gamma_{\rm S}$/kT&$\Gamma_{\rm H}$&$N_{\rm H}$&$\chi^{2}$/dof\\
\hline
{\bf IC5063}&po+po&2.16$^{+0.29}_{-0.31}$&1.70$^{+0.18}_{-0.19}$&2.15$\pm{0.18}$
$\times$10$^{23}$&260/263\\
\dotfill&brem+po&1.39$^{+1.02}_{-0.42}$&1.66$^{+0.20}_{-0.18}$&
2.07$^{+0.20}_{-0.19}$ $\times$10$^{23}$&262/263\\
\hline
{\bf NGC3998}&po&\dotfill&1.88$\pm{0.02}$&$\sim$$N_{{\rm H}_{\rm gal}}$
&1036/688\\
\hline
\end{tabular}
\end{table}

\section{NGC 3998}
NGC 3998 is a LINER galaxy (z=0.0035) with strong UV emission 
(Reichert et al. 1992), but weak X-ray luminosity 
($L_{\rm X}$$\sim$3.5$\times$10$^{41}$ ergs s$^{-1}$ in the 2-10 keV energy 
range). 
NGC 3998 was observed twice by ROSAT (May 91-May 92), for a total of $\sim$65 
Ks of observation, and once by ASCA (May 94), for which only GIS data were 
available (40 Ks/GIS).\\ 
The joint ASCA and ROSAT analysis gives a good fit with a power law model 
($\Gamma$ = 1.88$\pm{0.02}$) and a weak excess of absorption above the 
galactic value. An upper limit of 0.88 is obtained for the reflection 
component. 
There is also evidence 
of an iron line, whose parameters are given in table 2. 
No flux or spectral variability 
have been detected by ASCA and ROSAT. The low X-ray flux may be due to the 
fact that NGC 3998 is a low-luminosity AGN, but the high UV state suggests 
the presence of a massive black hole (M$\sim$10$^{8}$ M$_{\odot}$, 
Fabbiano et al. 1994), therefore in a steady-quiescent state, 
or a more elusive advection-dominated AGN (Fabian \& Rees 1995).\\
%
%
%

\centerline{\bf Tab. 2 - FeK$\alpha$ parameters}

\begin{table}[h]
\hspace{3.6cm} 
\begin{tabular}{|l|c|c|c|}
\hline
\hline
source&$E_{{\rm K}\alpha}$&$\sigma_{{\rm K}\alpha}$&EW\\
\hline
{\bf IC 5063}&6.35$^{+0.09}_{-0.11}$&$<$0.31&168$^{+138}_{-105}$\\
\hline
{\bf NGC 3998}&6.44$^{+0.16}_{-0.30}$&$<$0.41&200$^{+161}_{-139}$\\
\hline
\end{tabular}
\end{table}

\end{document}